\documentclass[twocolumn,showpacs,preprintnumbers,amsmath,amssymb,superscriptaddress]{revtex4}

\usepackage{graphicx}
\usepackage{graphicx,epsf} 
\usepackage{dcolumn}
\usepackage{bm}
\usepackage{latexsym}
\newcommand{\kav}{\langle k \rangle}
\newcommand{\xsize}{\epsfxsize=7.0cm}

\begin{document}

\author{Gerald~Paul}
\email{gerryp@bu.edu}

\author{Sameet Sreenivasan}

\author{H.~Eugene Stanley}

\affiliation{Center for Polymer Studies and Dept.\ of Physics, Boston
University}

\pacs{84.35.+i, 02.50.Cw, 05.50.+q, 64.60.Ak}
 
\title{Resilience of Complex Networks to  Random Breakdown}

\begin{abstract}

Using Monte Carlo simulations we calculate $f_c$, the fraction of nodes
which are randomly removed before global connectivity is lost, for
networks with scale-free and bimodal degree distributions.  Our results
differ with the results predicted by an equation for $f_c$ proposed by
Cohen, et al.  We discuss the reasons for this disagreement and clarify
the domain for which the proposed equation is valid.

\end{abstract}  

\maketitle

\section{Introduction}
It has been shown \cite{Molloy,Cohen2000,Callaway,Cohen2002} that random
uncorrelated networks with degree distribution $P(k)$ lose global
connectivity when
\begin{equation}
\kappa\equiv\frac{\langle k^2 \rangle}{\kav} < 2.
\label{kappa2}
\end{equation}
As explained in Ref.~\cite{Cohen2000,Cohen2002}, random removal of a
fraction $f$ of nodes from a network with degree distribution $P_0(k)$
results in a new degree distribution
\begin{equation}
P(k) = \sum_{k_0 = k}^K P_0(k_0) \binom{k_0}{k} \left( 1 - f \right)^k
f^{k_0 - k}.
\label{rand}
\end{equation}
Using this degree distribution to calculate $\kav$ and $\langle k^2
\rangle$ after random removal of sites it was determined
\cite{Cohen2000,Cohen2002} that
\begin{equation}
f_c=1-{1 \over{\kappa_0-1}}
\label{fc}
\end{equation}
where $\kappa_0$ is the value of $\kappa$ computed from the original
degree distribution, before the random removal. Equation (\ref{fc}) was
observed to hold for a number of network types, including random networks
that have a Poisson degree distribution, and was used in the analysis of
scale-free networks that have power-law degree distributions
\cite{Cohen2000,Cohen2002}.

Using Monte-Carlo simulations we find that Eq.~(\ref{fc}) does not hold for
networks with (i) self-loops and multiple edges and/or (ii) high
variance in $f_c$.  We illustrate our findings using scale-free and
bimodal networks and clarify the domains where Eq.~(\ref{fc}) is valid.

\section{Monte Carlo Simulations}

We create random networks having specified degree distributions using
the method described in Ref.~\cite{Molloy}.  We then randomly delete
nodes in the network and after each node is removed, we calculate
$\kappa$.  When $\kappa$ becomes less than 2 we record the number of
nodes $i$ removed up to that point.  This process is performed for
many realizations of random graphs with a specified degree distribution
and, for each graph, for many different realizations of the sequence of
random node removals. The threshold $f_c$ is defined as 
\begin{equation}
f_c \equiv \frac{ \langle i \rangle}{N}
\label{fcdef}
\end{equation}
where $\langle i \rangle$ is the average value of $i$.

\section{Scale-Free Networks}
\label{sf}

\begin{figure}
\centerline{
\xsize
\epsfxsize=6.0cm
\epsfclipon
\epsfbox{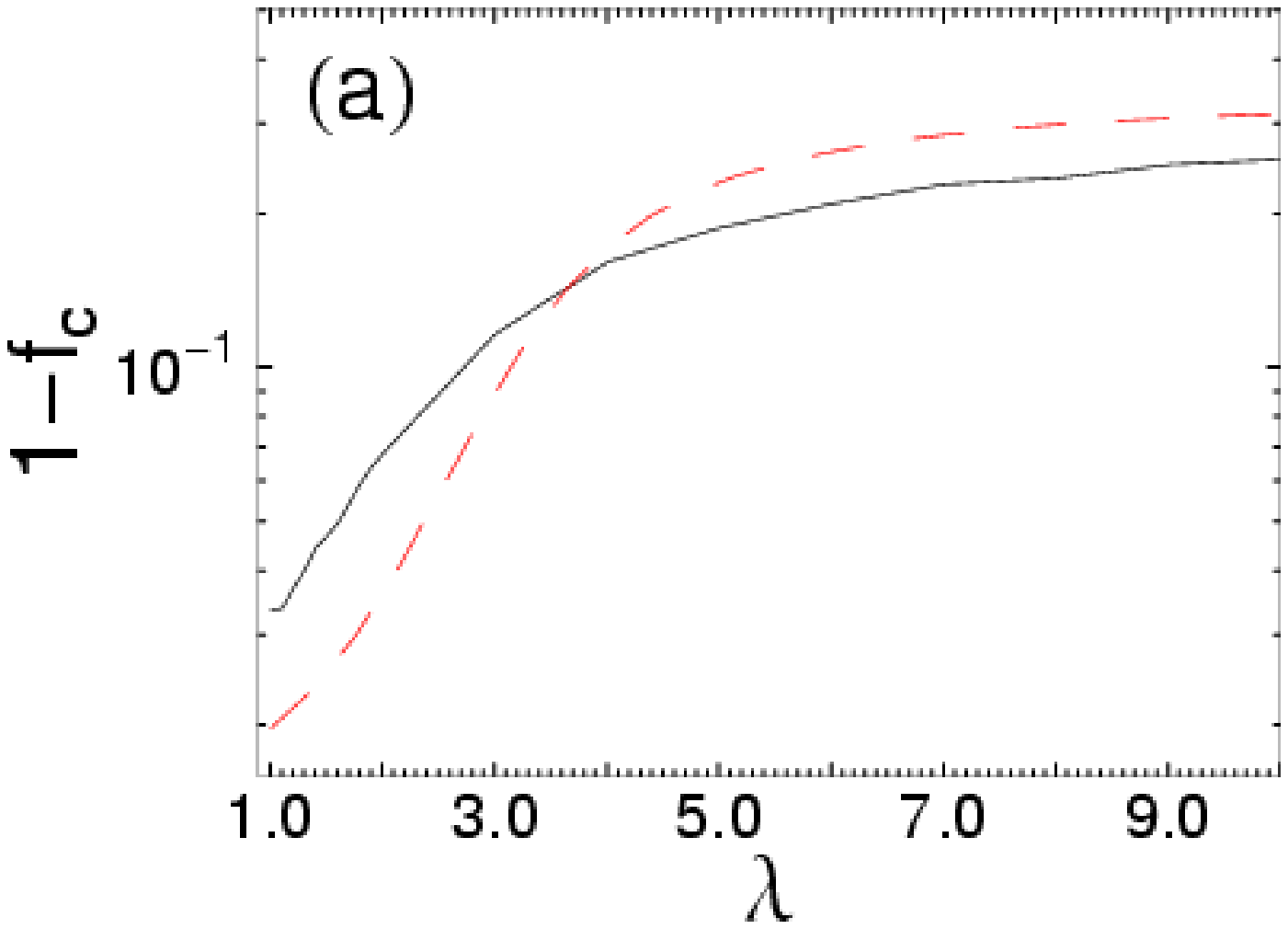}
}

\centerline{
\xsize
\epsfxsize=6.0cm
\epsfclipon
\epsfbox{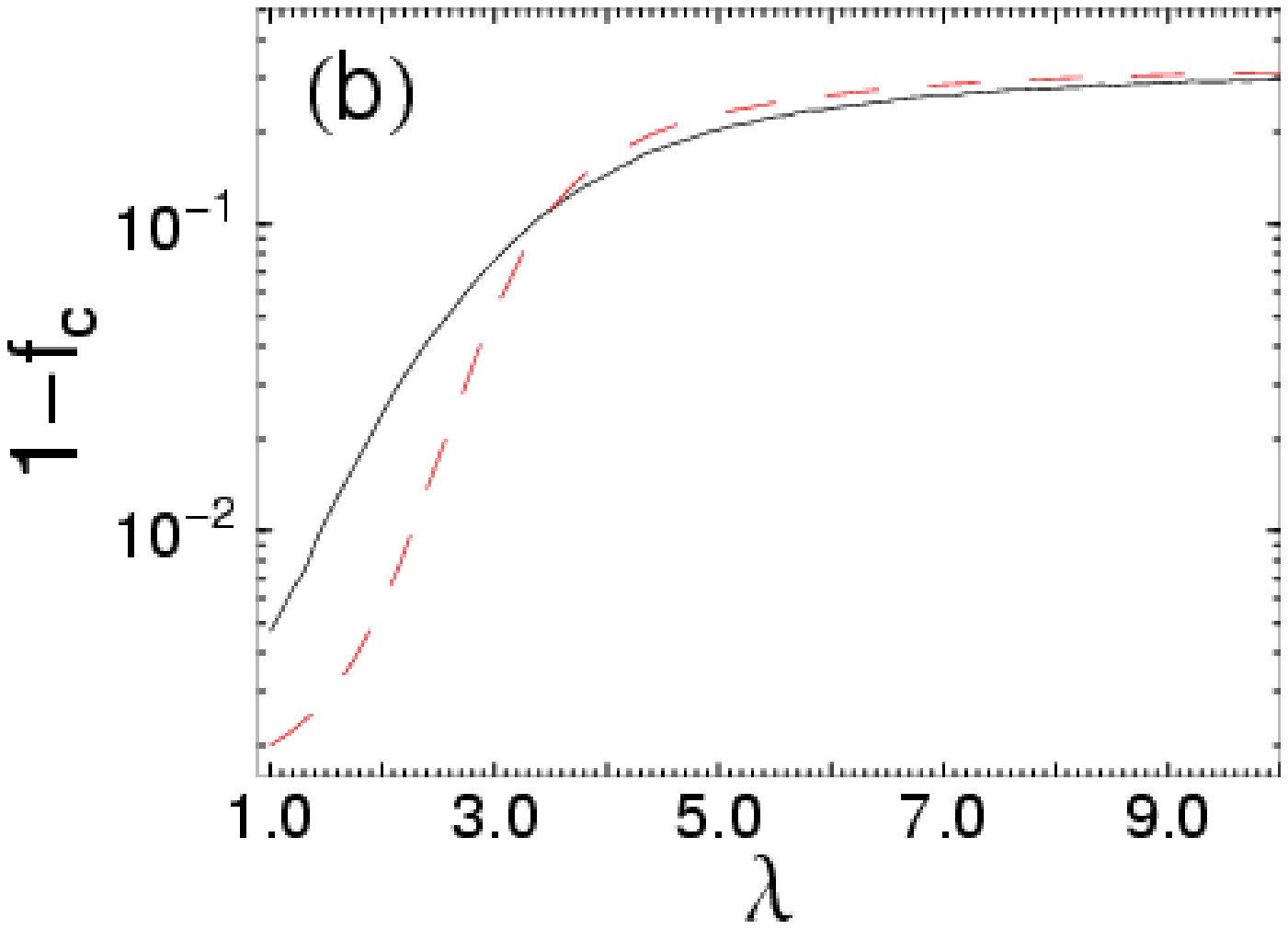}
}

\centerline{
\xsize
\epsfxsize=6.0cm
\epsfclipon
\epsfbox{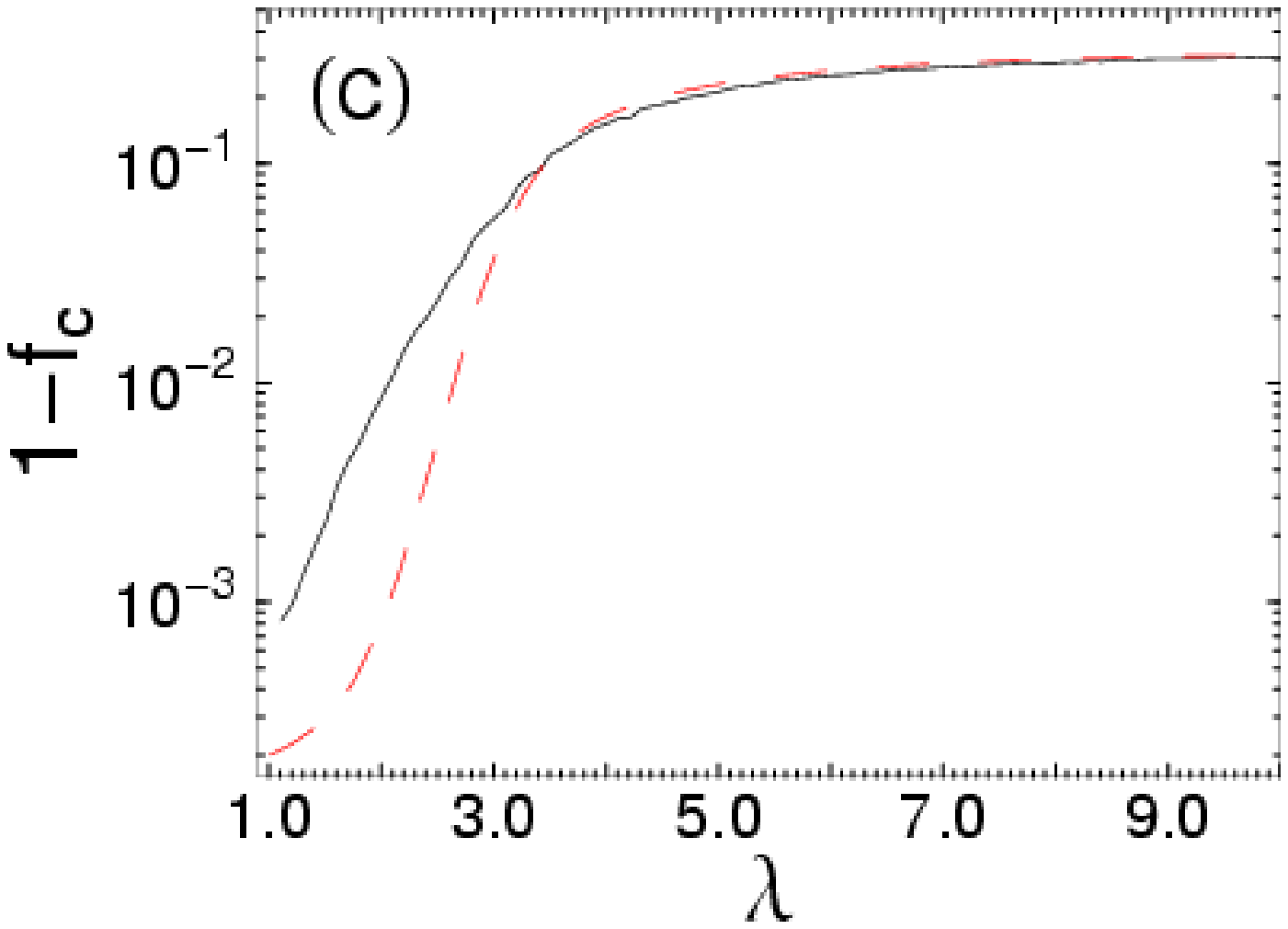}
}

\centerline{
\xsize
\epsfxsize=6.0cm
\epsfclipon
\epsfbox{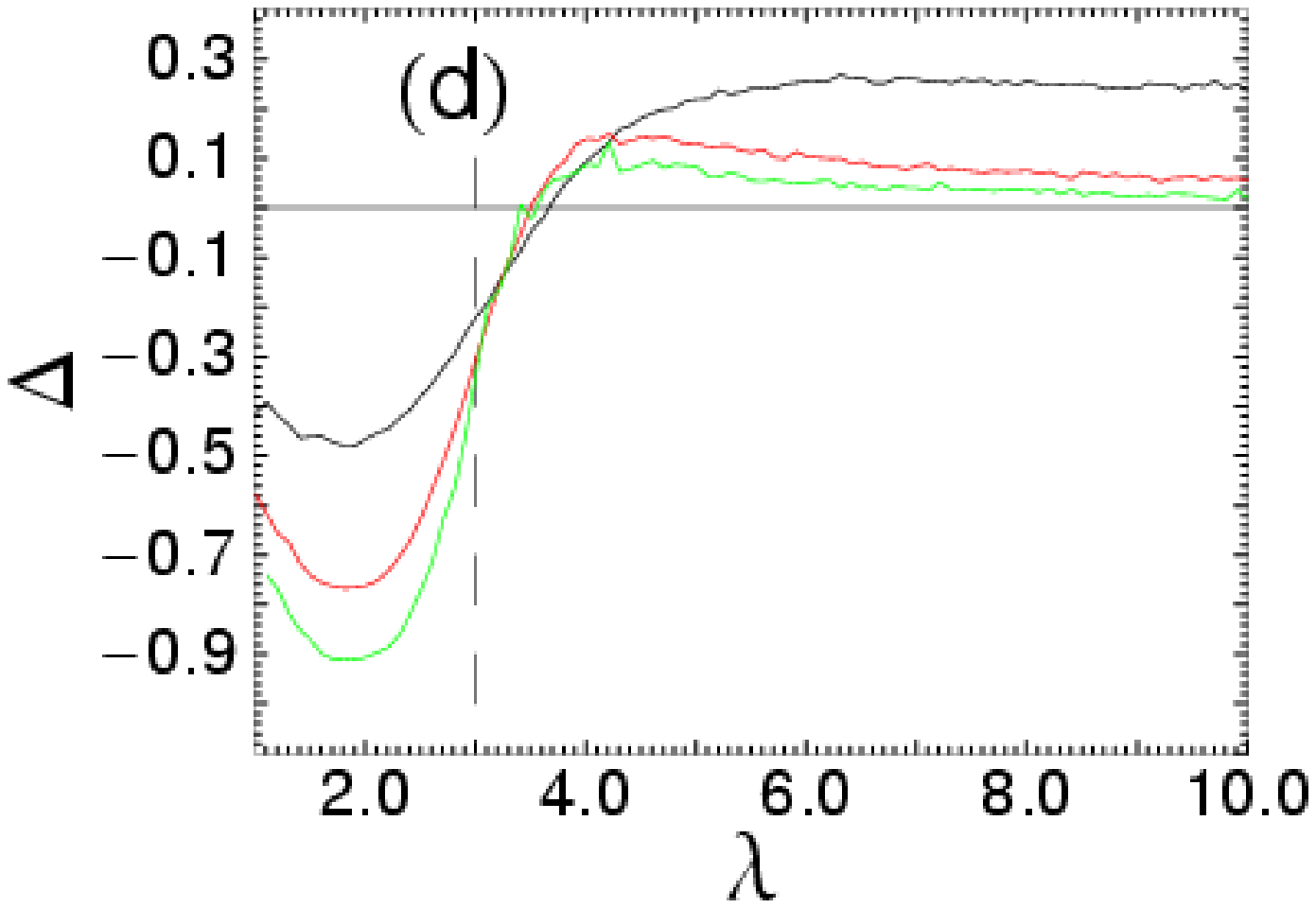}
}
\newpage
\caption{For $N=10^2, 10^3,$ and $10^4$ respectively in (a), (b) and (c),
  $1-f_c$ versus $\lambda$.  The solid line represents the results of
  Monte-Carlo simulations; the dashed line is the prediction of
  Eq.~(\ref{fc}).  (d) The difference $\Delta$ (see Eq.~(\ref{Delta}))
  between the prediction of Eq.~(\ref{fc}) and Monte-Carlo simulations
  for (from top to bottom) $N=10^2, 10^3, 10^4$. Note that if we
  had used a larger value of the upper cutoff $K$, then $\Delta$ would
  decrease monotonically from $\lambda=3$ to $\lambda=1$ instead of
  having a minimum near $\lambda=2$.}
\label{pk}
\end{figure}

We study scale-free random networks with degree distribution 
\begin{equation}
P(k)\sim k^{-\lambda} \qquad\qquad [m \le k \le K].
\end{equation}
We choose the lower cutoff $m=4$ and the upper cutoff $K=N$.  In
Figs.~\ref{pk}(a), (b) and (c), we show the dependence on $\lambda$ of
$1-f_c^{\rm MC}$ obtained by the Monte Carlo simulations and compare it
with $1-f_c^{\rm th}$ obtained theoretically from Eq.~(\ref{fc}).  The
simulation results agree well with Eq.~(\ref{fc}) for
$\lambda>\lambda^\ast$, where $\lambda^\ast\approx 3$, and the agreement
becomes better for increasing $N$.  However, for $\lambda <
\lambda^\ast$ there is significant disagreement, and the disagreement
becomes larger as $N$ increases, as seen clearly Fig.~\ref{pk} (d) in
which we plot the normalized difference
\begin{equation}
\Delta \equiv { f_c^{\rm th} - f_c^{\rm MC} \over{f_c^{\rm MC}} }.
\label{Delta}
\end{equation} 

\begin{figure}
\centerline{
\xsize
\epsfclipon
\epsfbox{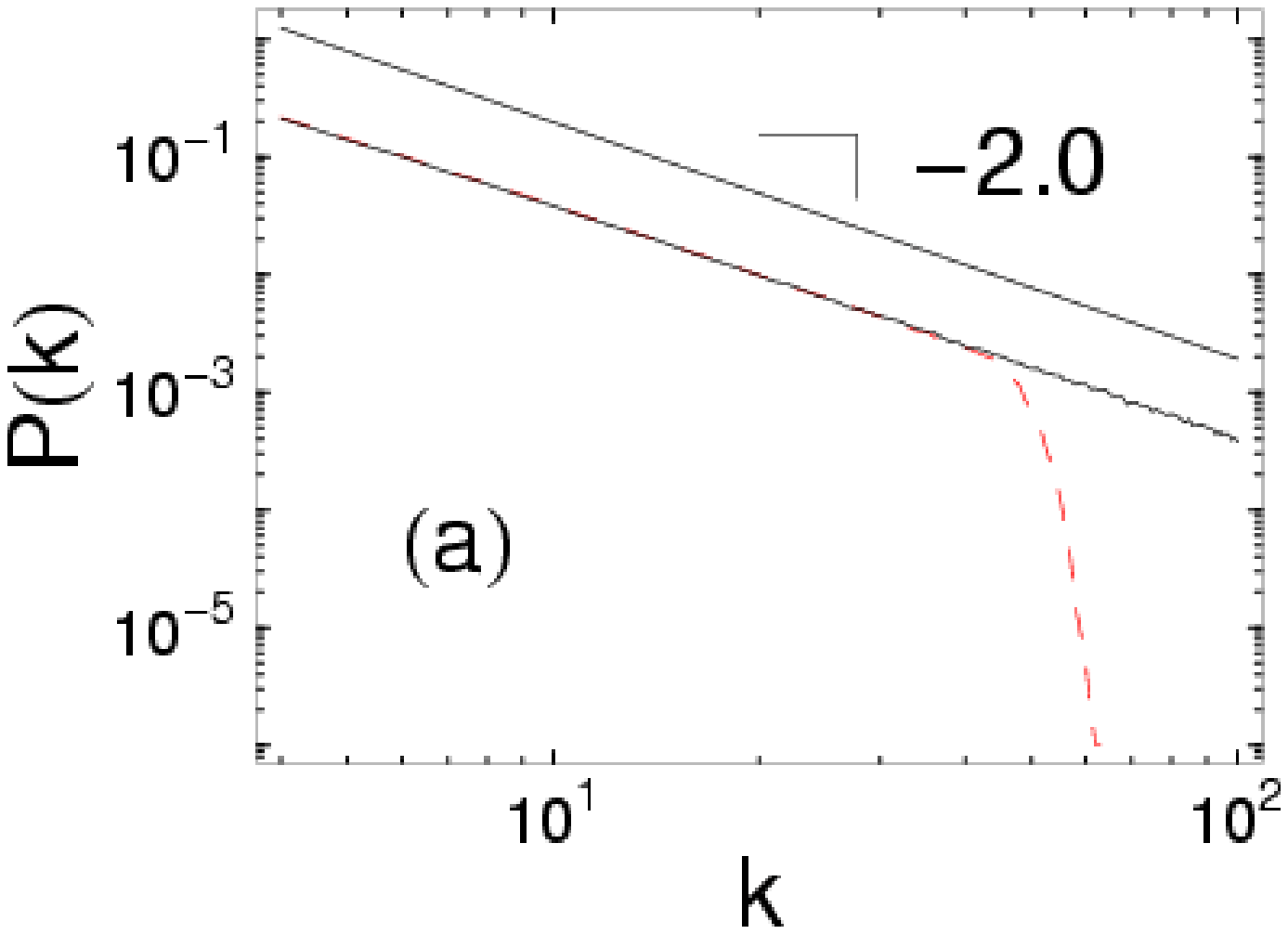}
}

\centerline{
\xsize
\epsfclipon
\epsfbox{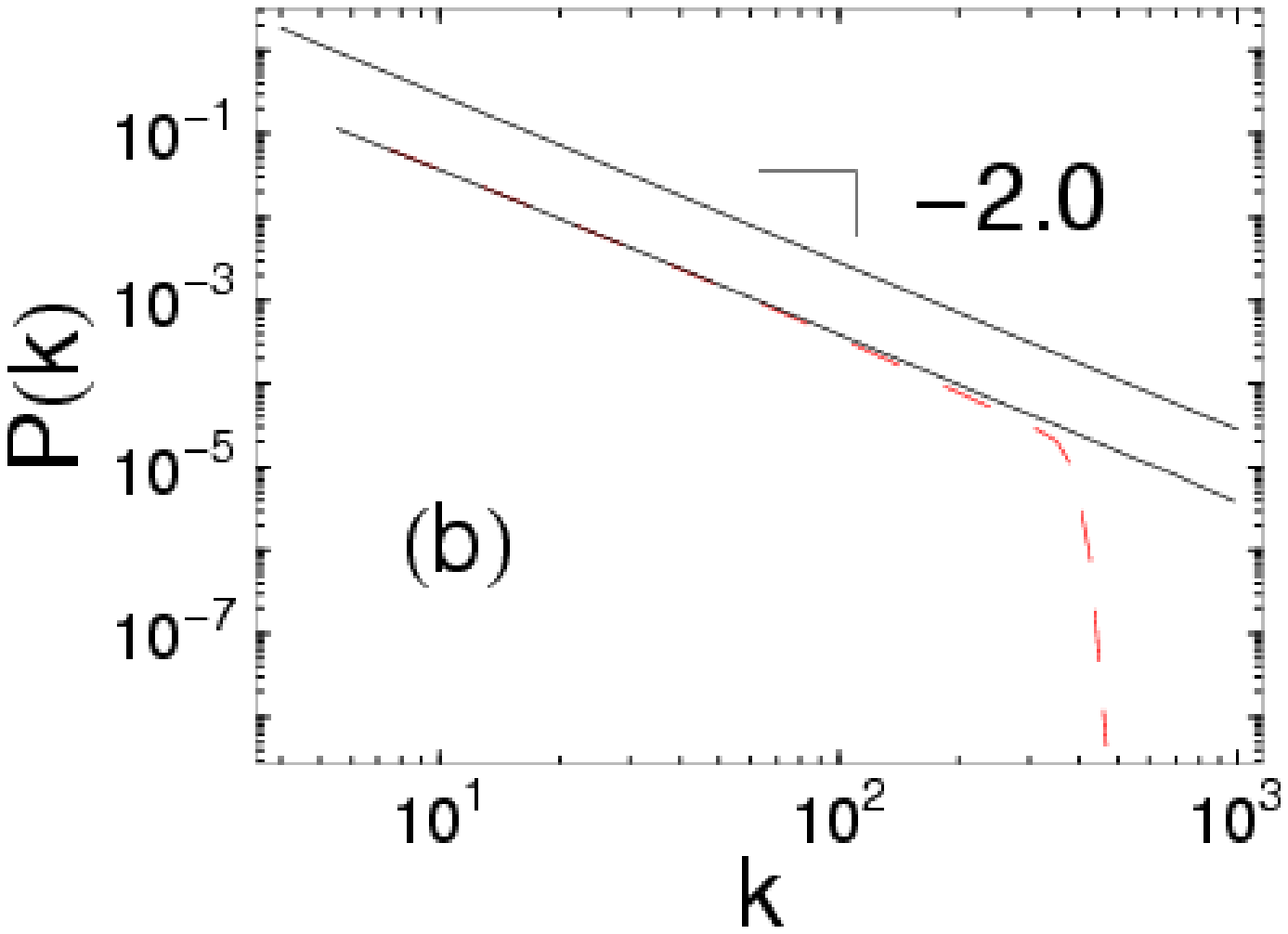}
}

\centerline{
\xsize
\epsfclipon
\epsfbox{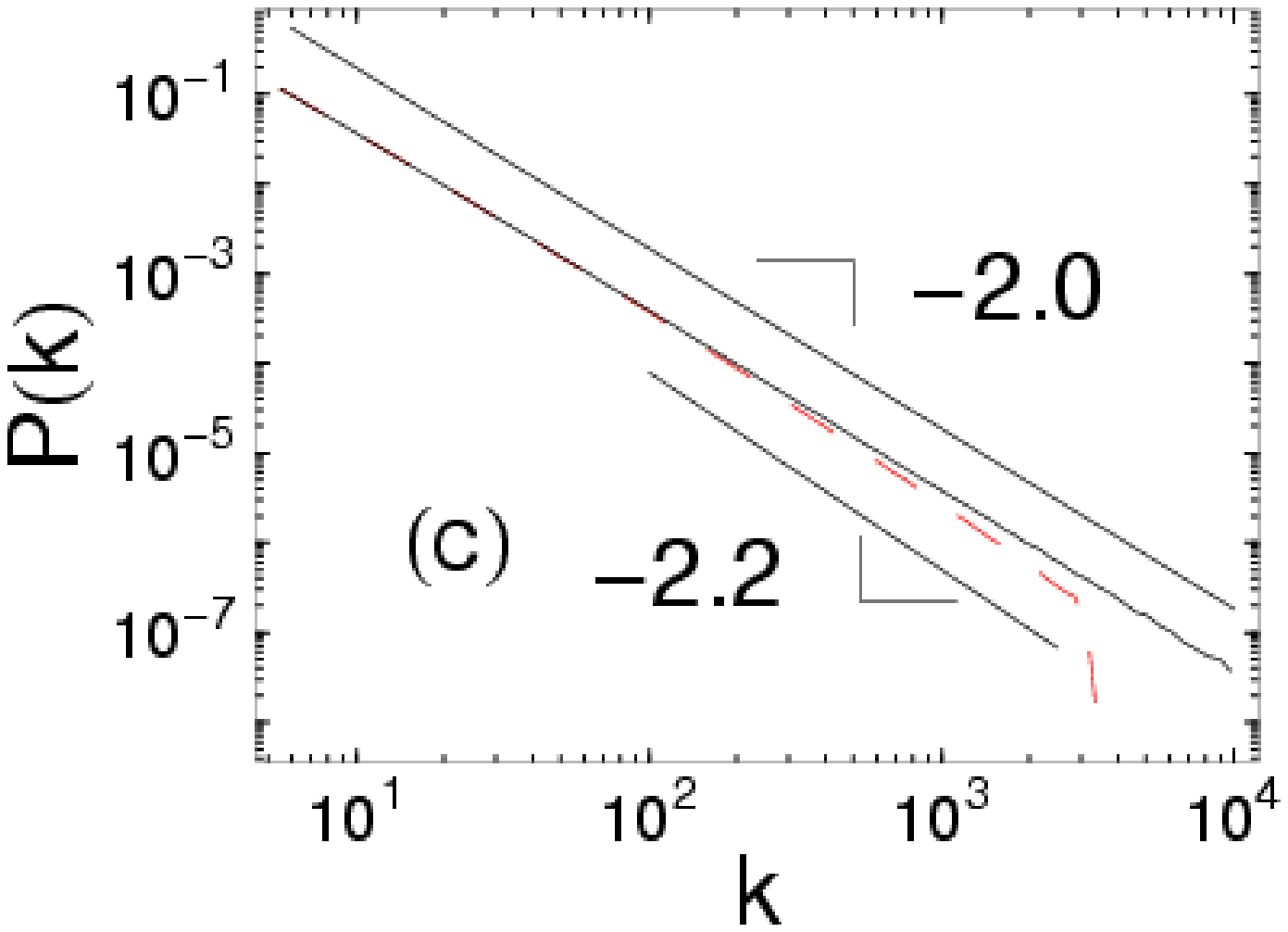}
}
\caption{$P(k)$ versus $k$ for $N=10^2, 10^3, 10^4$ in (a),(b) and (c)
  respectively.  The solid line represents $P(k)$ after network
  construction using the Molloy-Reed method; the dashed line is the
  distribution after the removal of self-loops and multiple edges. }
\label{pdist}
\end{figure}

\begin{figure}
\centerline{
\xsize
\epsfclipon
\epsfbox{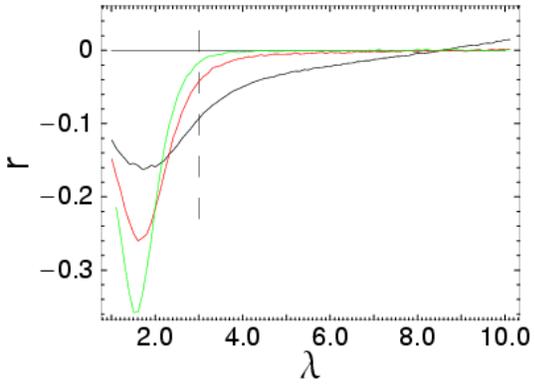}
}

\caption{Correlation $r$ as a function of $\lambda$ for (from top to
  bottom at left) $N=10^2, 10^3$, and $10^4$ for distributions after
  removal of self-loops and multiple edges.  Note that the correlation
  increases with $N$ for $\lambda \protect \lesssim 3$ and decreases
  with $N$ for $\lambda \protect \gtrsim 3$}.
\label{pcorr}
\end{figure}

The nonzero value of $\Delta$ has its root in the use of
Eq.~(\ref{rand}) to derive Eq.~(\ref{fc}).  Equation~(\ref{rand}) is
valid only if, in the original network, two conditions hold: (i) There
are no {\it self loops}, i.e.  all links from node $i$ are to distinct
nodes $j$ with $j \ne i$ and (ii) there are no multiple links between
$i$ and $j$. In graph theory networks satisfying these two conditions
are called {\it simple}.  If the original network is not simple,
Eq.~(\ref{rand}) must then be interpreted as operating on the original
network but with self-loops and multiple links deleted.  But this
deletion changes the properties of the degree distribution. As seen in
Figs.~\ref{pdist} (a), (b), and (c) the cutoff is changed, and for large
N, the slope of the tail of the distribution is modified.  Also the
degrees of adjacent nodes become correlated as seen in Fig.~\ref{pcorr},
which shows the $\lambda$-dependence of the degree correlation
\cite{Newman2002}
\begin{equation}
 r \equiv { {1\over \sigma_q^2}} \sum_{j,k}(e_{jk} -q_j q_k).
\label{r}
\end{equation}
Here $e_{jk}$ is the joint probability of the remaining degrees \cite{note1}
 of the two vertices at either end of a randomly chosen
edge, $q_k$ is the probability of the remaining degree of a single
vertex at the end of a randomly chosen edge, and
\begin{equation}
\sigma_q^2 \equiv \sum k^2 q_k -\left(\sum_k k q_k\right)^2.
\label{sigmaq}
\end{equation}
Because of the degree correlations, Eq.~(\ref{kappa2}) no longer applies
and therefore Eq.~(\ref{fc}) no longer holds.  The similarity in
appearance between Fig.~\ref{pk}(d) and Fig.~\ref{pcorr} confirms that
the nonzero correlations play a major role in the difference between
$f_c^{\rm MC}$ and $f_c^{\rm th}$.

We can explain the domain of validity of Eq.~(\ref{fc}) as follows. It
is known \cite{Chung,Burda,Boguna,Catanzaro} that for any desired random
degree distribution, the networks created by such methods as those of
Molloy-Reed \cite{Molloy} or Chun-Lu \cite{Chung} create simple graphs
only if $P(k)=0$ for $k$ greater than the {\it structural cutoff}
\begin{equation}
K_s \equiv \sqrt{\kav N}.
\label{KcStruct}
\end{equation}
It is also known that for scale-free networks the number of nodes with
degree greater than the {\it natural cutoff}
\begin{equation}
K_c \equiv m N^{1/(\lambda-1)} 
\label{KcNat}
\end{equation}
is statistically insignificant \cite{Cohen2000,Dorogovstev2002}.  These
two facts are sufficient to understand that Eq.~(\ref{fc}) is valid for
scale-free networks only if $\lambda > 3$ (in which case the natural
cutoff $K_c$ results in nodes with degree $ \gtrsim \sqrt{N}$ being
statistically insignificant) or for $\lambda < 3 $ if the maximum degree
is less than the structural cutoff $K_s$.


\section{Bimodal Networks}
\label{bimodal}
\subsection{Star Networks}

First, we discuss a simple example with a bimodal degree distribution
for which Eq.~(\ref{fc}) fails.  Consider a {\it star network} of N nodes with
degree distribution
\label{pkxx}
\begin{equation}
\label{Pk1}
P(k)=\begin{cases}
(N-1)/N & [k=1]\\
    1/N & [k=N-1]
\end{cases}
\end{equation}
and $P(k)=0$ for all other values of $k$. If nodes are randomly removed,
the criterion for losing global connectivity, $\kappa < 2$, is obtained
when the single node with degree $N-1$, the {\it hub} node, is removed
or when almost all of the degree 1 nodes, the {\it leaf} nodes, are
removed. The probability that almost all the leaf nodes are removed
before the hub node is removed approaches 0 for large $N$.  Let $i$ be
the number of nodes which are removed before the hub node is
removed. Since the removal is random, $i$ is uniformly distributed
between $0$ and $N-1$ and, from Eq.~(\ref{fcdef}), $f_c=1/2$.  On the
other hand, Eq.~(\ref{fc}) predicts $f_c=1-2/N$ which asymptotically
approaches unity for large $N$.

\begin{figure}
\centerline{
\xsize
\epsfclipon
\epsfbox{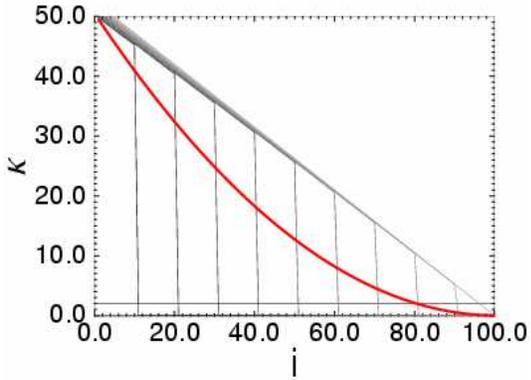}
} 
\caption{Example illustrating case in which $\langle (i|\kappa=2 \rangle
 \ne (i | \langle \kappa \rangle =2)$ for star network of 1 hub of
 degree 99 and 99 nodes of degree 1.  Thin lines are $\kappa$ vs $i$,
 where $i$ denotes the number of the step at which a node is deleted, for
 cases in which the hub is deleted at step (from left to right)  1, 10,
 20, 30, 40, 50, 60, 70, 80, 90 and 100.  The thick line is the average
 of the thin lines.  Note that the value of $i$ at which the average
 crosses the horizontal line $\kappa=2$ is much higher than $50$, the
 average of the values of $i$ at which the thin lines cross the
 horizontal line $\kappa=2$. }
\label{pav}
\end{figure}

As for the case of scale-free networks, we can understand this
disagreement as a result of the presence of self loops. We can also use
this star network example to identify another implicit assumption used in the
derivation of Eq.~(\ref{fc}), namely that
\begin{equation}
 \langle i\rangle \equiv   \langle (i|\kappa(i)=2)\rangle=(i|\langle\kappa(i)
\rangle=2)
\label{ne}
\end{equation}
where $\kappa(i)$ is the value of $\kappa$ after the removal of $i$
nodes. That is, we define $\langle i \rangle$ to be the
average of $i$ such that in each random removal $\kappa(i)=2$; the
derivation of Eq.~(\ref{fc}) assumes that $\langle i \rangle$ is equal
to $i$ such that the average of $\kappa(i)$ over all random removals
equals 2. Equation~(\ref{ne}) will be true in the limit in which the
variance $\langle(i-\langle i\rangle)^2\rangle$ is zero.  But when the
variance becomes large as is the case for the star network,
Eq.~(\ref{ne}) may be not hold.  Figure~\ref{pav} illustrates
graphically an example for which Eq.~(\ref{ne}) does not hold because
the variance in $i$ is large.

\subsection{Other Bimodal Networks}

\begin{figure}
\centerline{
\xsize
\epsfclipon
\epsfbox{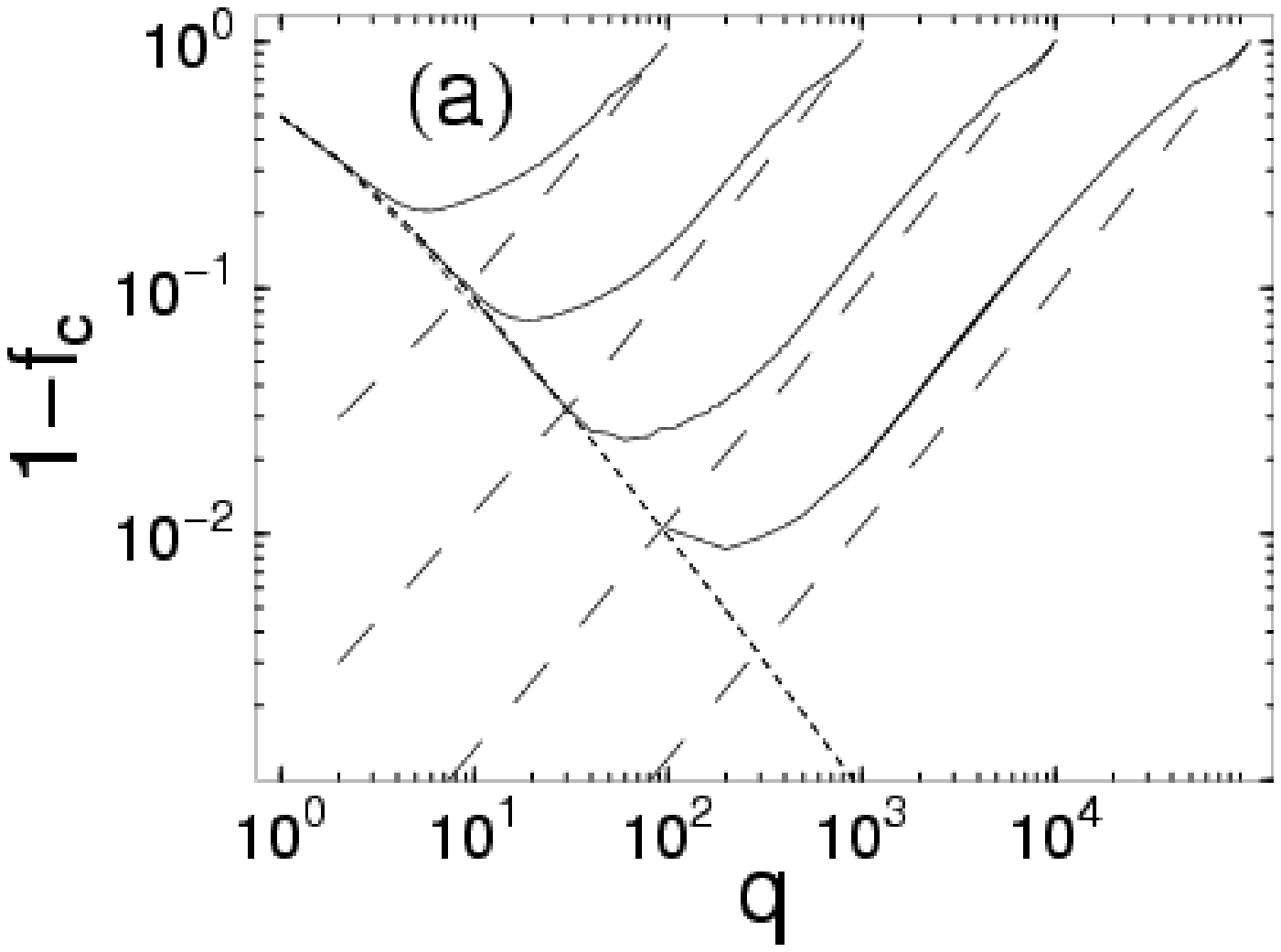}
}

\centerline{
\xsize
\epsfclipon
\epsfbox{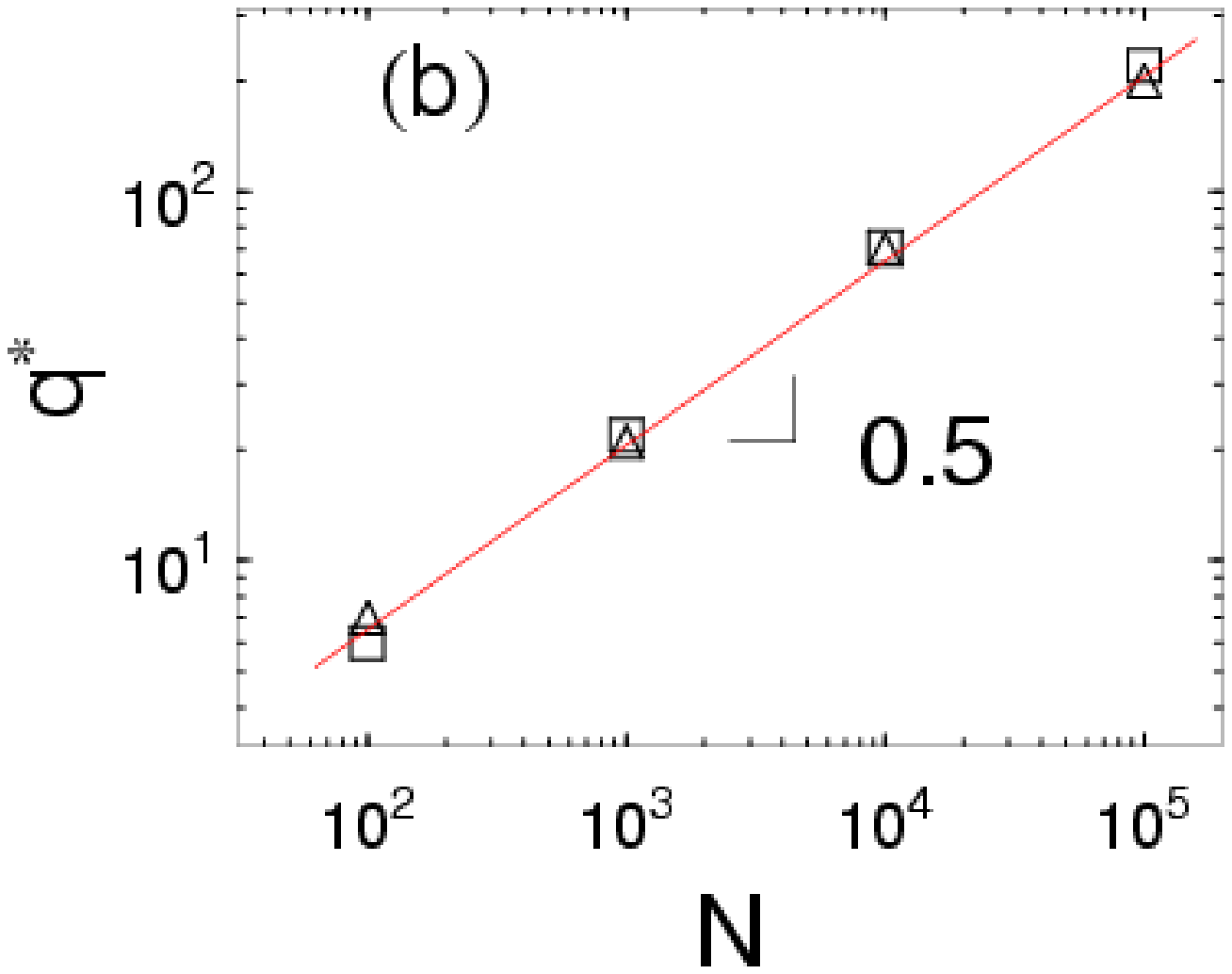}
}

\caption{For $\kav=2$ and for (from left to right) $N=10^2,10^3,10^4$ and
 $10^5$ (a) $1-f_c$ vs. number of hubs $q$.  The solid lines represent
 Monte-Carlo simulation results.  Dashed lines(short) are approximation
 $f_c^{\rm low}$; dashed lines(long) are approximation $f_c^{\rm high}$.
 (b)Number of hubs, $q$ versus $N$.  Squares represent characteristic
 values $q^*$ at which high and low $q$ approximations
 intersect. Triangles represent values of $q$ at which the standard
 deviation in $1-f_c$ is minimal.}
\label{kav2}
\end{figure}

In order to study other bimodal networks, we extend the star network to
networks with $q$ high degree hubs connected to the remaining nodes of
degree one.  For networks with average degree $\kav$, the degree
distribution is specified as
\begin{equation}
\label{pkx}
P(k)=\begin{cases}
(N-q)/N & [k=1]\\
    q/N & [k=k_2]
\end{cases}
\end{equation}
where
\begin{equation}
k_2={(\kav-1) N + q \over{q}},
\label{k2}
\end{equation}
and $P(k)=0$ for all other $k$.  We first consider networks with
$\kav=2$.  In Fig.~\ref{kav2}(a), for the distribution of
Eqs.~(\ref{pkx}) and (\ref{k2}), we plot $1-f_c$ as a function of $q$
for $N=10^2$, $10^3$, $10^4$, and $10^5$.  Also shown in
Fig.~\ref{kav2}(a) are plots for approximations $f_c^{\rm high}$ and
$f_c^{\rm low}$ which we expect to be valid respectively for high and
low values of $q$. We will use these approximations to determine how
$f_c(q)$ scales and for which values of $q$ Eq.~(\ref{fc}) is valid.  The
approximations are determined as follows:

\begin{itemize}

\item[{(i)}] When $q \sim N$, (i.e., the network is homogeneous) we expect
Eq.~(\ref{fc}) to hold so $f_c^{\rm high}=1-1/(\kappa_0-1)$.

\item[{(ii)}] For small $q$, the network loses global connectivity when
all $q$ high degree nodes are removed.  The probability that all $q$
high degree nodes are removed after the first $i$ nodes of all types
have been removed is
\begin{equation}
g(q,N,i)={ q \over{N}} { {i-1 \choose{q-1}}  \over{{N-1 \choose{q-1}} }}.
\label{g}
\end{equation}

\end{itemize}
Here $i$ is now the average number of nodes that must be removed before all
$q$ high degree nodes are removed. Then $ \langle i\rangle =\sum_{i=q}^N
i g(q,N,i)$ and
\begin{equation}
\label{fclow}
f_c^{\rm low}={\langle i\rangle  \over{N}}={\sum_{i=q}^N i g(q,N,i) \over {N}}.
\end{equation}
Note that $f_c^{\rm low}$ does not depend on $\kav$ since changing
$\kav$ results simply in a different number of links between the high
degree nodes; if our criterion for collapse is the removal of all high
degree nodes, the number of links between them is irrelevant.  As
expected, the plots of $f_c^{\rm low}$ and $f_c^{\rm high}$ approximate
the values of $f_c$ for low and high values of $q$, respectively.

\begin{figure}
\centerline{
\xsize
\epsfclipon
\epsfbox{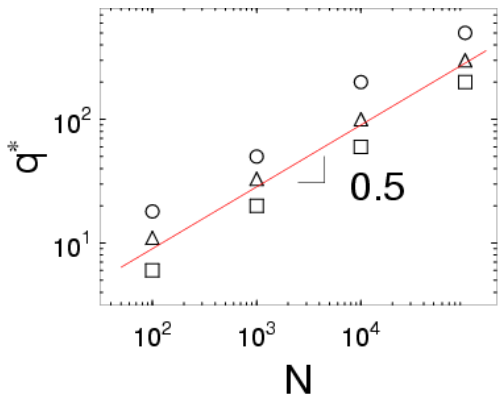}
}
 \caption{(a) Number of hubs, $q^*$, at which approximations for low and
  high $q$ intersect vs. $N$.  Squares, triangles and circles represent
  networks with $\kav=2,3,$ and 4 respectively.}
\label{comb234}
\end{figure}

In Fig.~\ref{kav2}(b), we plot the the number of hubs, $q^*$, for which the
functions $f_c^{\rm low}(q)$ and $f_c^{\rm high}(q)$
intersect. We find that
\begin{eqnarray}
q^* \sim N^{0.5}          
\label{qfintersect}
\end{eqnarray}
Similar plots (see Fig.~\ref{comb234}) for $\kav=3$ and $\kav=4$ also
exhibit scaling of $q^*$ as $N^{0.5}$ with only a change in the
prefactor; the scaling is independent of $\kav$.

The simulation results suggest that $q^*$ scales as $\sqrt{N}$.  We can
show this to be the case by solving analytically for $q^*$ for large $N$
as follows: For general $\kav$, using the distribution in
Eqs.~(\ref{pkx}), we find for $N \gg q \gg 1$
\begin{equation}
f_c^{\rm high}=1-{q \over{(\kav-1)N}}.
\label{fchigh}
\end{equation}
For $f_c^{\rm low}$, the sum in Eq.~(\ref{fclow}) can be performed
analytically, yielding
\begin{equation}
\label{fclow1}
f_c^{\rm low}={\Gamma(N+2)(\Gamma(q+2)-\Gamma(q+1))  \over{N \Gamma(N+1)
    \Gamma(q+2)}} 
\end{equation}
for $q >0$.  For large N, 
\begin{equation}
\label{fclow2}
f_c^{\rm low}={\Gamma(q+2)-\Gamma(q+1)  \over{\Gamma(q+2)}}.
\end{equation}
To first order in $1/q$, Eq.~(\ref{fclow2}) yields
\begin{equation}
\label{fclow3}
f_c^{\rm low}=1-{1 \over{q}} + O({1 \over{q^2}}).
\end{equation}
Equating Eqs.~(\ref{fclow3}) and (\ref{fchigh}) we find
\begin{equation}
\label{qintersect}
q*=\sqrt{\kav-1} \sqrt{N}
\end{equation}
consistent with the plot in Fig.~\ref{kav2}(b) and
Eq.~(\ref{qfintersect}).  From the fact that $q*$ scales like
$\sqrt{N}$, we conclude that all characteristic values of $f_c$ scale
like $\sqrt{N}$ with a prefactor dependent on $\kav$. In particular the
value of $q$ at which $f_c^{\rm MC}$ (found from Monte Carlo
simulations) agrees to any desired degree with the value of $f_c^{\rm
th}$ (from Eq.~(\ref{fc})) will scale with $N$ in the same fashion
in which $q^*$ scales with $N$, Eq.~(\ref{qfintersect}).  For
simplicity, we consider Eq.(3) to be valid for $q > q^*$.

\begin{figure}
\centerline{
\xsize
\epsfclipon
\epsfbox{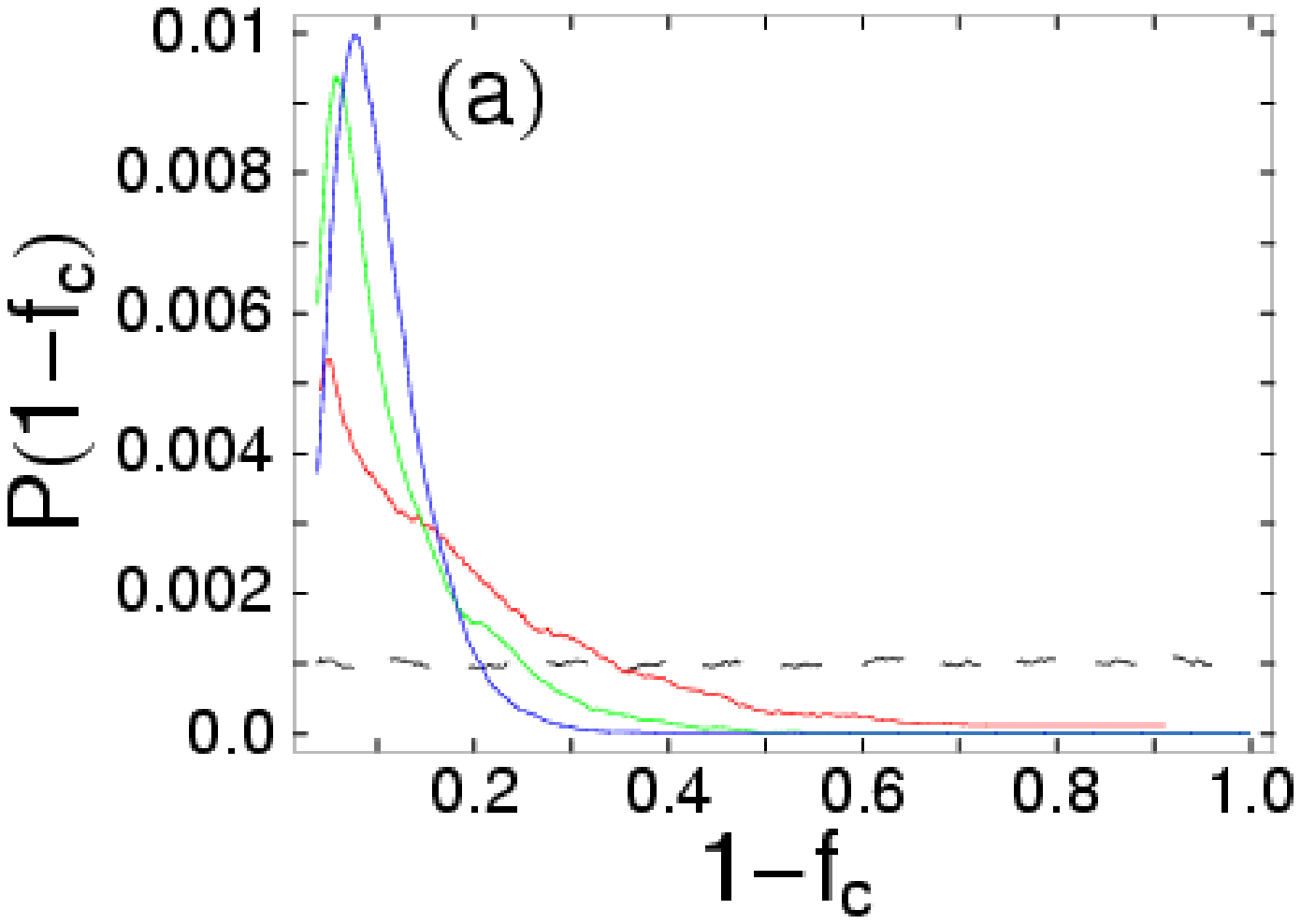}
}

\centerline{
\xsize
\epsfclipon
\epsfbox{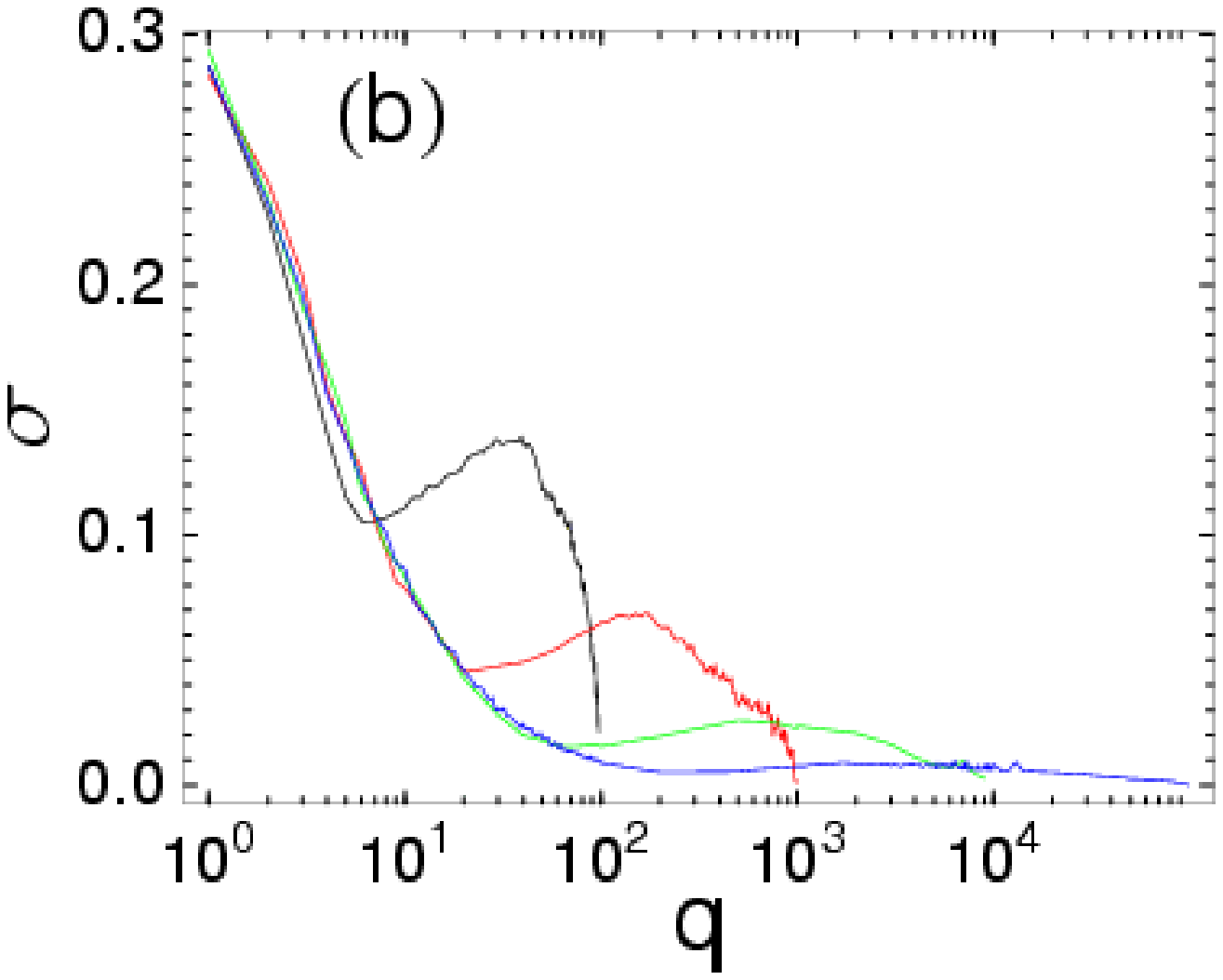}
}
 \caption{(a) $P(1-f_c)$ the probability distribution of $1-f_c$ for
 $N=10^3$ and $q=1$(dashed line) and (from left to right in order of
 increasing position of peaks) $q=5,10,$ and $20$.  (b) Standard
 deviation $\sigma$ versus $q$ for $N=10^2,10^3,10^4$ and $10^5$
 (from left to right in order of increasing length of the tails of the
 distributions). Note that the second peak in this plot which is most
 pronounced for smaller $N$ is an artifact of finite size}.
\label{dev}
\end{figure}

We now confirm that the variance in $f_c$ is in fact small
for values of $q$ for which Eq.~(\ref{ne}) holds.  In Fig.~\ref{dev}(a),
for $N=10^3$ and $q=1$, 5, 10, and 20, we plot $P(1-f_c)$ vs. $1-f_c$.
As expected, for $q=1$ (star network) the distribution is uniform
because there is an equal probability that the single high degree node
will be removed at any value of $i$.  For the larger values of $q$, the
distributions $P(1-f_c)$ develop a well-defined peak.  To quantify the
definition of these peaks, we plot in Fig.~\ref{dev}(b), the standard
deviation of $f_c$
\begin{equation} 
\sigma={\sqrt{\langle i^2 \rangle - \langle i \rangle ^2} \over{N}}
\label{sigma}
\end{equation}
versus $q$ for $N=10^2$, $10^3$, $10^4$, and $10^5$.  Each of the plots
has a large deviation at $q=1$ and decrease to a local minimum, the
position of which $\tilde q$ increases with increasing $N$.  For $q$
greater than the $\tilde q$, the deviation is small and decreases with
increasing $N$.  In Fig.~\ref{kav2}(b) we plot $\tilde q$ as a function
of $N$. We see that the values of these minima are essentially the same
as the values of $q^*$, the value of $q$ above which Eq.~(\ref{fc}) is
valid.  This is consistent with our understanding that Eq.~(\ref{fc}) is
valid when the variance is small.

\subsection{Domain of Validity}

\begin{figure}
\centerline{
\xsize
\epsfclipon
\epsfbox{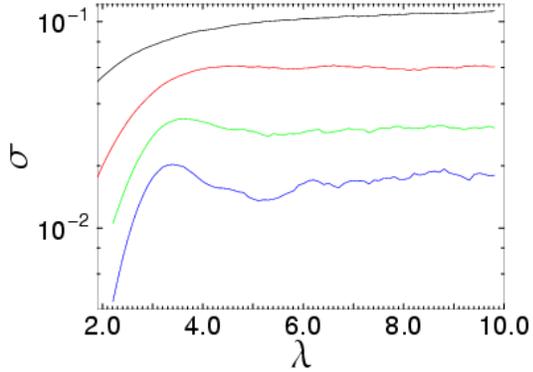}
}
 \caption{For random scale-free networks with $4 \le k \le N$, standard
deviation $\sigma_{fc}$ versus $\lambda$ for $N=10^2,10^3,10^4$ and
$10^5$ (from top to bottom).}
\label{devsf}
\end{figure}

Since $q$ and the degree of the hubs $k_2$ are related by
Eq.~(\ref{k2}), we can determine for what values of $k_2$ Eq.~(\ref{ne})
is valid. Substituting Eq.~(\ref{qintersect}) in Eq.~(\ref{k2}) we find
that Eq.~(\ref{ne}) is valid when
\begin{equation}
k_2  < \sqrt{(\kav-1)N}.
\label{k2max}
\end{equation}
Thus the criterion for Eq.~(\ref{ne}) holding is essentially the same as
the criterion discussed in Sec.~\ref{sf} for the graph being simple.
The bimodal networks we study here in which a relatively small number of
nodes control the global connectivity of the network yield large
variances in $f_c$ for networks with a given number of nodes; in fact,
for $q=1$ the worst case variance is obtained.  This suggests that the
criterion of Eq.~(\ref{k2max}) may hold for all degree distributions as
a requirement for a low variance in $f_c$. If this is the case, we can
use the requirement that $P(k) =0$ for $k \lesssim K_c$ as the criterion
for both the network being simple and $f_c$ having a small variance.
Note, however, that while the criteria are similar, it is not true that
the presence of self-loops and multiple edges implies that the
distribution of $f_c$ has a large variance; for example, the variance of
$f_c$ in scale-free networks is small even in the presence of self-loops
and multiple edges, as seen in Fig. ~\ref{devsf}.

\section{Discussion and Summary}
 
We have clarified the domain of validity of Eq.~(\ref{fc}), a general
equation for determining $f_c$, the fraction of nodes which must be
randomly removed before global connectivity is lost.  For  Eq.~(\ref{fc})
to be valid, (i) the highest degree of any nodes present in
statistically significant numbers in a random network must be less than
the structural cutoff $K_s \equiv \sqrt{\kav N}$ and (ii) the variance
of $f_c$ must be small.  For bimodal networks the variance in $f_c$ is
small when the hubs have degree less than $\sqrt{(\kav-1) N}$. That the
bimodal networks we have studied represent a worst case for large
variance suggests that in general the criterion that the network be
simple is sufficient for Eq.~(\ref{ne}) to hold.  It is not clear if
there is a deeper connection between these two criteria both of which scale
as $\sqrt{N}$.

\subsubsection*{Acknowledgments}

We thank S. Havlin for helpful discussions and ONR for support.


\begin{thebibliography}{99}
\bibitem{note2} The notation $(x|y)$ should be read as ``x such that
condition y holds''.
\end{thebibliography}
\end{document}